# Thermal Spin-Orbit Torque in Spintronics

Zheng-Chuan Wang

The University of Chinese Academy of Sciences, P. O. Box 4588, Beijing 100049, China.

**Abstract**

Within the spinor Boltzmann equation (SBE) formalism, we derived a temperature dependent thermal spin-orbit torque based on local equilibrium assumption in a system with Rashba spin-orbit interaction. If we expand the distribution function of spinor Boltzmann equation around local equilibrium distribution, we can obtain the spin diffusion equation from SBE, then the spin transfer torque, spin orbit torque as well as thermal spin-orbit torque we seek to can be read out from this equation. It exhibits that this thermal spin-orbit torque originates from the temperature gradient of local equilibrium distribution function, which is explicit and straightforward than previous works. Finally, we illustrate them by an example of spin-polarized transport through a ferromagnet with Rashba spin-orbit coupling, in which those torques driven whatever by temperature gradient or bias are manifested quantitatively.

**PACS**: 72.00, 65.90.+i, 85.70.-w, 85.80.-b

# I. Introduction

In last decades, spintronics provide an alternative way to manipulate the local magnetic moment of ferromagnet in the mesoscopic devices by the magnetization switching resulting from spin transfer torque (STT)[1], in particular in the magnetoresistance random-access memory (MRAM). However, the spin transfer torque can only be used to reverse the magnetization of noncollinear system, i.e., the Ferromagnet/ Insulator/Ferromagnet tunneling junction, where the two ferromagnetic layers have different directions of magnetization and need an additional polarizing layer. Recently, another spin torque--spin-orbit torque (SOT) had been proposed[2-6], which can be employed to switch the magnetization in a collinear system with a broken inversion symmetry, i.e., in the system of racetrack memory (RM). Since only a low critical current density is needed to drive the fast domain wall motion in RM, SOT had attracted more and more attentions[7-10]. Till now, a new branch of spintronics —spin-orbitronics had been established.

Similar to the thermal spin transfer torque given by Hatami et al.[11], the SOT can also be driven by the temperature gradient instead of voltage bias, which is called thermal spin-orbit torque (TSOT). The TSOT was originally presented by Freimuth et al. by Kubo's linear response formalism, they expressed the even torque in terms of

Berry phase[12-13]. Besides this TSOT initiated from the spin-polarized electronic current, the magnon-mediated TSOT was also predicted theoretically by Manchon et al. in 2014[14-15]. Nowadays, the interplay between the SOT and thermoelectric transport had been intensively explored whatever in theories or experiments[16-20]. However, Freimuth's Berry phase expression of TSOT need a knowledge of electronic structure of magnetic system from the first principle calculation, it is time-consuming in some complicated system, so in this manuscript, we try to give another explicit expression for TSOT by means of distribution function.

The main focus of our work is to investigate the TSOT by the spinor Boltzmann equation (SBE) which was accomplished by Levy et al. in 2002[21], Sheng et al. ever proposed a similar equation at steady state in 1997[22]. At present, SBE had become a powerful tool to explore the spin-polarized transport in spintronics. In addition to calculate the magneto-resistance in the electronic transport, SBE can also be used to study the STT with respect to magnetization switching, Zhang et al. ever generalized the STT to the non-adiabatic case by means of SBE[23]. SBE was also extended to the case beyond gradient approximation[24]. Subsequently, Wang et al. included the Rashba and dresslhaus spin-orbit coupling into the SBE[25], which is helpful for us to find the SOT from spin diffusion equation derived from SBE. In this manuscript, we will show that a generalized SOT can be singled out from the spin diffusion equation, which contains not only the usual SOT, but also a term coming from the temperature gradient, it is just the TSOT we want.

## II. Theoretical Formalism

In spintronics, the SBE with Rashba spin-orbit coupling in a magneto-electric system under an external electric field $\vec{E}$ had been given by us in 2019[25],

$$\left(\frac{\partial}{\partial t}+\frac{\vec{p}}{m}\cdot\vec{\nabla}-e\vec{E}\cdot\vec{\nabla}_p\right)\hat{f}(p,x,t)+\frac{iJ}{\hbar}[\vec{M}\cdot\vec{\sigma},\hat{f}(p,x,t)]+\frac{\alpha}{2\hbar}\{(\vec{\nabla}\times\vec{z})\cdot\vec{\sigma},\hat{f}(p,x,t)\}$$

$$+\frac{i\alpha}{2\hbar^2}[(\vec{p}\times\vec{z})\cdot\vec{\sigma},\hat{f}(p,x,t)]=-\left(\frac{\partial\hat{f}}{\partial t}\right)_{collision}, \qquad (1)$$

where $\hat{f}(p,x,t)$ is the spinor distribution function which is a $2\times 2$ matrix, $J$ is the exchange coupling constant, $\vec{M}(x)$ is the unit vector of magnetization in the ferromagnet, $\vec{\sigma}$ denotes the Pauli matrix vector. $\alpha(\vec{\nabla}\times\vec{z})\cdot\vec{\sigma}$ describes the Rashba spin-orbit interaction which is induced by the structure inversion asymmetry, $\alpha$ is the coupling constant, $\vec{z}$ is the direction of symmetry axis in crystal. $\left(\frac{\partial\hat{f}}{\partial t}\right)_{collision}$ represents the collision term. Under the local equilibrium assumption, the spinor distribution function can be decomposed as[21]

$$\hat{f}(p,x,t)=f^0(p,x)+\left(-\frac{\partial f^0}{\partial\varepsilon}\right)[f(p,x,t)+\vec{g}(p,x,t)\cdot\vec{\sigma}], \qquad (2)$$

where $f^0(p,x)=\frac{1}{\exp\left[\frac{\varepsilon-\mu}{kT(x)}\right]+1}$ is the local equilibrium distribution, $\mu$ is the chemical potential. The temperature $T(x)$ is position-dependent and its gradient will give rise to the thermal spin current in the magnetic system, $k$ is the Boltzmann constant. The scalar distribution function $f(p,x,t)$ and vector distribution function $\vec{g}(p,x,t)$ in Eq.(2) describe the deviation of spinor Boltzmann function from the local equilibrium.

Substituting Eq.(2) into the SBE (1), under the relaxation approximation we can divide the SBE into the equation for the scalar distribution function as

$$\left(\frac{\partial}{\partial t}+\frac{\vec{p}}{m}\cdot\vec{\nabla}-e\vec{E}\cdot\vec{\nabla}_p\right)f_0(p,x)-\left(\frac{\partial}{\partial t}+\frac{\vec{p}}{m}\cdot\vec{\nabla}-e\vec{E}\cdot\vec{\nabla}_p\right)\left(\frac{\partial f_0}{\partial \varepsilon}f\right)-\frac{\alpha}{\hbar}(\vec{\nabla}\times\vec{z})\left(\frac{\partial f_0}{\partial \varepsilon}\cdot\vec{g}(p,x,t)\right)=\frac{-\frac{\partial f_0}{\partial \varepsilon}f}{\tau}, \qquad (3)$$

and the equation for the vector distribution function as

$$-\left(\frac{\partial}{\partial t}+\frac{\vec{p}}{m}\cdot\vec{\nabla}-e\vec{E}\cdot\vec{\nabla}_p\right)\left(\frac{\partial f_0}{\partial \varepsilon}\vec{g}(p,x,t)\right)+\frac{J}{\hbar}\frac{\partial f_0}{\partial \varepsilon}\vec{M}\times\vec{g}(p,x,t)+\frac{\alpha}{\hbar}(\vec{\nabla}\times\vec{z})f_0(p,x)-\frac{\alpha}{\hbar}(\vec{\nabla}\times\vec{z})\left(\frac{\partial f_0}{\partial \varepsilon}f\right)+\frac{\alpha}{\hbar^2}(\vec{p}\times\vec{z})\times\left(\frac{\partial f_0}{\partial \varepsilon}\vec{g}(p,x,t)\right)=\frac{-\frac{\partial f_0}{\partial \varepsilon}\vec{g}(p,x,t)+<\frac{\partial f_0}{\partial \varepsilon}\vec{g}>}{\tau_{sf}}, \qquad (4)$$

they are coupled together. The scalar and vector distribution functions obtained may enable us to calculate those physical observables in the spin-polarized transport. Usually, we define the charge density and charge current density by the scalar distribution function as $\rho(x,t)=e\int\frac{\partial f_0}{\partial \varepsilon}f(p,x,t)dp$ and $\vec{j}(x,t)=e\int\vec{v}\frac{\partial f_0}{\partial \varepsilon}f(p,x,t)dp$, respectively, and define the spin accumulation and spin current density by the vector distribution as $\vec{m}(x,t)=e\int\frac{\partial f_0}{\partial \varepsilon}\vec{g}(p,x,t)dp$ and $\vec{j}_m(x,t)=e\int\vec{v}\frac{\partial f_0}{\partial \varepsilon}\vec{g}(p,x,t)dp$, respectively, similarly the thermal current density can also be defined as $\vec{j}_E(x,t)=\int\varepsilon\vec{v}\frac{\partial f_0}{\partial \varepsilon}f(p,x,t)dp$. So we can calculate these physical observables after we have solved the scalar and vector distribution functions from Eqs. (3) and (4).

If we integrate the momentum $p$ over the Fermi surface on the both sides of Eq. (3), we have,

$$\frac{\partial}{\partial t}\rho(x,t) + \vec{\nabla}\cdot\vec{j}(x,t) = -\int\left(\frac{\partial}{\partial t} + \frac{\vec{p}}{m}\cdot\vec{\nabla} - e\vec{E}\cdot\vec{\nabla}_p\right)f_0(p,x)dp + \frac{\alpha}{\hbar}\int(\vec{\nabla}\times\vec{z})\cdot(\frac{\partial f_0}{\partial \varepsilon}\vec{g}(p,x,t))dp, \tag{5}$$

which is the continuity equation for charge density and charge current defined above. Analogously, if we integrate the momentum $p$ on the both sides of Eq. (4), we have

$$\frac{\partial}{\partial t}\vec{m}(x,t) + \vec{\nabla}\cdot\vec{j}_m(x,t) = -\frac{J(p)}{\hbar}\left(\vec{M}(x)\times\vec{m}(x)\right) - \frac{\alpha}{\hbar}\int(\vec{\nabla}\times\vec{z})f_0(p,x)dp + \frac{\alpha}{\hbar}(\vec{\nabla}\times\vec{z})\int\left(\frac{\partial f_0}{\partial \varepsilon}f\right)dp - \frac{\alpha}{\hbar^2}\int(\vec{p}\times\vec{z})\times(\frac{\partial f_0}{\partial \varepsilon}\vec{g}(p,x,t))dp, \tag{6}$$

where the spin current is a tensor. This is the continuity equation for the spin accumulation and spin current. Eq.(6) is also called the spin diffusion equation, which is critical for us to read out the STT from this equation[24-25].

After the spin-flip relaxation time $t \gg \tau_{sf}$, the system will arrive at a steady state, where $\frac{\partial}{\partial t}\vec{m}(x,t) = 0$, then the spin diffusion equation (6) will reduce to

$$\frac{J(p)}{\hbar}\left(\vec{M}(x)\times\vec{m}(x)\right) = -\vec{\nabla}\cdot\vec{j}_m(x,t) - \frac{\alpha}{\hbar}(\vec{\nabla}\times\vec{z})\int f_0(p,x)dp + \frac{\alpha}{\hbar}(\vec{\nabla}\times\vec{z})\int\left(\frac{\partial f_0}{\partial \varepsilon}f\right)dp - \frac{\alpha}{\hbar^2}\int(\vec{p}\times\vec{z})\times(\frac{\partial f_0}{\partial \varepsilon}\vec{g}(p,x))dp, \tag{7}$$

where we have moved the spin torque $\frac{J(p)}{\hbar}\left(\vec{M}(x)\times\vec{m}(x)\right)$[24-25] to the left hand side and the divergence of spin current $\vec{\nabla}\cdot\vec{j}_m(x,t)$ to the right hand side of Eq.(7). As a consequence of this rearrangement, we can see that the spin torque are contributed by several parts: the divergence of spin current which is concerned with the usual STT[23]; the temperature dependent term $-\frac{\alpha}{\hbar}(\vec{\nabla} \times \vec{z}) \int f_0(p,x)dp$, we refer to this as the thermal SOT, because it is produced by the gradient of local equilibrium distribution function, namely the temperature gradient, so this term is just the thermal SOT we looking for, it is the central result in our manuscript. The fourth and the fifth terms on the right hand of Eq.(7) correspond to the usual SOT, we ever derived them in Ref.[25] . Our expression of TSOT is explicit, which is convenient for us to calculate it in some complex systems. In the next, we will evaluate these torques numerically.

III. Numerical Results

As an example, we consider a spin-polarized transport through a ferromagnet under an external electric field with a broken inversion symmetry, there exists Rashba spin-orbit coupling in this system. If we keep a temperature distribution in the system as $T(x) = T_0 + kx$, where $T_0$ is a constant, $k$ is the temperature gradient, $x$ denotes the position, then this temperature gradient will induce a thermal spin current, which will produce the TSOT.

In order to quantify those torques and currents defined in the above section, we should solve Eq.(3) combining with (4) simultaneously, because they are the coupled differential equations about the scalar distribution function and vector distribution function. Since only the electrons near the Fermi surface mainly contribute to the transport, we can firstly approximate the $e\vec{E} \cdot \vec{\nabla}_p(\frac{\partial f_0}{\partial \varepsilon} f)$ in the lefe hand of Eq.(3) as $e\vec{E} \cdot \vec{\nabla}_p f_0$ [24,25], and $e\vec{E} \cdot \vec{\nabla}_p(\frac{\partial f_0}{\partial \varepsilon} \vec{g}(p,x))$ in the left hand of Eq.(4) as $e\vec{E} \cdot \vec{\nabla}_p \vec{g}^0(p,x)$, where $\vec{g}^0(p,x)$ is a approximated solution at steady state, $\vec{g}^0(p,x) = (\exp\left[i\frac{J(M_y - M_z)}{\hbar v}x\right], \exp\left[i\frac{J(M_z - M_x)}{\hbar v}x\right], \exp\left[i\frac{J(M_x - M_y)}{\hbar v}x\right])$, which is analogous to the form given by Levy et al.[21], these approximations can be improved by the iteration procedures in the latter step by step. These coupled equations may be solved by the method of Fourier transformation. After Fourier transform Eqs.(3) and (4) on both sides of them, we have

$$i\omega F(p,k,\omega) + i\vec{k} \cdot \frac{\vec{p}}{m} F(p,k,\omega) - \frac{F(p,k,\omega)}{\tau} + \frac{i\alpha}{\hbar}(\vec{k} \times \vec{z}) \cdot \vec{G}(p,k,\omega) = \int e^{i\omega t} e^{ikx}[-eE\frac{\partial}{\partial p}f_0(p,x) - \left(\frac{\partial}{\partial t} + \frac{\vec{p}}{m} \cdot \vec{\nabla} - e\vec{E} \cdot \vec{\nabla}_p\right)f_0(p,x) + \frac{f^0(p)}{\tau}]dxdt,$$

(8)

and

$$i\omega \vec{G}(p,k,\omega) + i\vec{k} \cdot \frac{\vec{p}}{m}\vec{G}(p,k,\omega) - \frac{J}{\hbar}\left(\vec{M} \times \vec{G}(p,k,\omega)\right) - \frac{\vec{G}(p,k,\omega)}{\tau} + \frac{i\alpha}{\hbar}(\vec{k} \times \vec{z})F(p,k,\omega) - \frac{\alpha}{\hbar^2}(\vec{p} \times \vec{z}) \times \vec{G}(p,k,\omega) = \int e^{i\omega t} e^{ikx}[-eE\frac{\partial}{\partial p}\vec{g}^0(p,x) + \frac{\alpha}{\hbar}(\vec{\nabla} \times \vec{z})f_0(p,x) + \frac{\vec{g}^0(p,x)}{\tau_{sf}}]dxdt,$$

(9)

where $F(p,k,\omega)$ and $\vec{G}(p,k,\omega)$ are the Fourier transformations of scalar distribution function $\frac{\partial f_0}{\partial \varepsilon} f$ and vector distribution function $\frac{\partial f_0}{\partial \varepsilon}\vec{g}(p,x)$, respectively. The advantage of Fourier transformation lies in the fact that Eqs.(8) and (9) are the linear equations group about $F(p,k,\omega)$ and $\vec{G}(p,k,\omega)$, which can be easily performed by some algebra operations, then making the inverse Fourier transformation, we can obtain the scalar distribution function and vector distribution function. After this, we further substitute these distribution functions obtained into the right hand of Eqs.(8) and (9) to improve the approximations $\vec{\nabla}_p f_0$ and $\vec{\nabla}_p \vec{g}^0(p,x)$ at first step, then solving Eqs.(8) and (9) again to get a new solutions, repeat this iteration procedure again and again until the iterated solutions converge. Finally we can evaluate the physical observables defined in the above by these converged distribution functions.

In our calculation, we apply an external electric field $\vec{E} = 36\mu V/nm$ to the system[13], the length of ferromagnet is chosen as $20nm$ and the coupling constant $J(p)$ for the magnetization of background ferromagnet as $1.2ev$[24]. Meanwhile, we adopt the momentum relaxation time as $\tau = 0.01ps$ and the spin-flip relaxation time as $\tau_{sf} = 0.1ps$[24]. The temperature distribution in the system is kept as $T(x) = T_0 + kx$, where $T_0 = 300K$, $k = 2K/nm$[13], we adopt the spin-orbit coupling constant as $\alpha = 1.0 \times 10^{-11} eV$, which is a typical value in some system, i.e., the two-dimensional electron gas[25]. The average $\langle f \rangle$ in Eq.(3) can be regarded as the local equilibrium distribution function $f^0(p,x)$, in this regard, $\langle \vec{g} \rangle$ in Eq.(4) can be approximated as $\vec{g}^0(p,x)$.

In Fig.1, we plot the charge current density as a function of position x and y in the case of steady state. We can see that the charge current density has a slowing variation in the system, the difference of it's maximum and minimum is small, this variation can be regarded as arising from the Rashba spin-orbit interaction in the system. The current is always positive because we apply a uniform electric field and temperature gradient to the system. In fact, the variation of current with respect to position and time is governed by the continuity equation (5), here we only give out the current at steady state. Except the charge current, there also exists thermal current. The thermal current density as a function of position is shown in

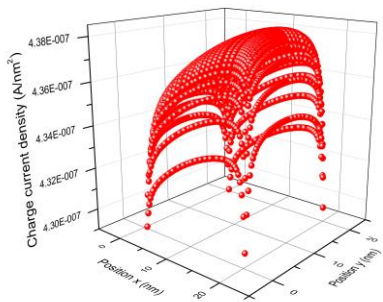 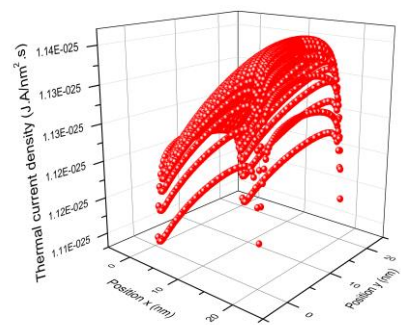

Fig.1 The charge current density vs position, where $T_0 = 300K, E = 36\mu V/nm$.

Fig.2 The thermal current density vs position, where $T_0 = 300K$, $E = 36\mu V/nm$.

Fig.2, it is similar to the charge current, which can be interpreted by its definition. The thermal current is produced not only by the temperature gradient but also by the electric field, which correspond to the first and second terms in the left hand side of Eq.(3), respectively. Here we only consider the thermal current carried by the conduction electrons, not by the phonons. In Fig.3, we draw the x-component of spin current density as a function of position x and y, it oscillates rapidly. It's y- and z- components have similar shapes with x-component, so we haven't shown them here. Since the spin current is a tensor, we only plot the component of spin current along x-axis. The position variation of spin current will give rise to the spin accumulation, both of them obey the spin diffusion equation (6). The spin accumulation will result in STT on the background ferromagnet. The numerical results for STT are illustrated in Fig.4, which is calculated by the expression $\frac{J(p)}{\hbar}\left(\vec{M}(x) \times \vec{m}(x)\right)$[24-25], so it is analogous to the spin accumulation. In a uniform magnetization, it is proportional to the spin accumulation. In order to address the contributions on the spin torque from the several terms in Eq.(7), we plot the usual SOT and thermal SOT in Fig.5 and Fig.6, respectively. It is obvious that the TSOT is smaller than the usual SOT, the SOT provides a major contribution to the total spin torque. As we know the SOT is driven by the voltage bias, while the TSOT is driven by the temperature gradient, they have different origins. If we regard the SOT and the TSOT as a unified spin-orbit torque, then the unified SOT is mainly dominated by the usual SOT. For further studying the characteristic features of TSOT, we plot the TSOT at different temperatures $T_0 = 300K, 200K, 100K$ in Fig.6, respectively. It is shown that the TSOT has a relative big magnitude at low temperature $T_0 = 100K$, so the TSOT will play an important role at the low temperature, we can not ignore it in this case. It should be noted that our SOT and TSOT only correspond to the odd torque in Freimuth et al's work[12-13], because it originates from the Rashba-like effect, but our expression is very explicit, it need not perform the first principle calculation, so it is helpful for us to more practical applications.

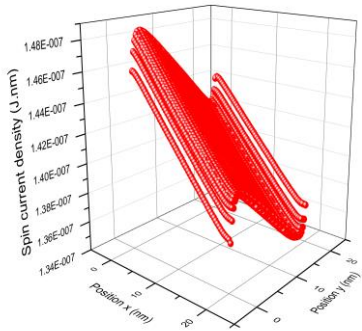
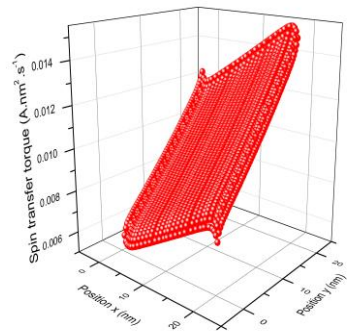

Fig.3 The x-component of spin current density vs position, where $T_0 = 300K, E = 36\mu V/nm$.

Fig.4 The y-component of STT vs position, where $E = 36\mu V/nm$.

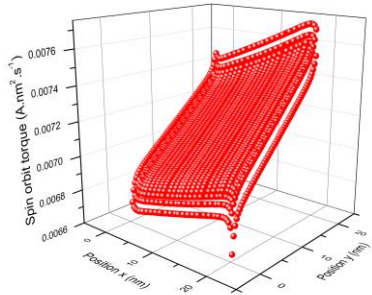
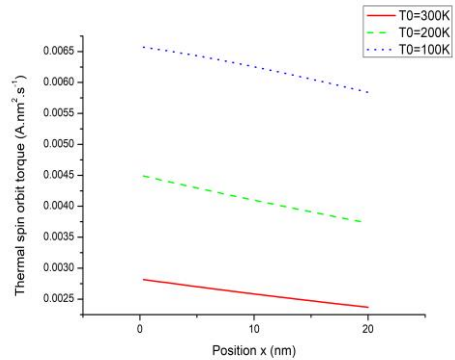

Fig.5 The y-component of SOT vs position, where $T_0 = 300K, E = 36\mu V/nm$.

Fig.6 The y-component of TSOT vs position at different temperature $T_0 = 100K, 200K, 300K$, where $E = 36\mu V/nm$.

IV. Summary and Discussions

Under the local equilibrium assumption, we derived a unified SOT in the framework of SBE, which consists of not only the usual SOT, but also the temperature dependent thermal SOT. The TSOT and the usual SOT have different origins, the former is induced by the temperature gradient, while the latter is induced by the voltage bias. The essence of our work lies in the fact that we express the TSOT by the local equilibrium distribution function instead of the Berry phase given by Freimuth[12-13], which is helpful for us to calculate it conveniently. The numerical results indicate that the TSOT is smaller than SOT, but it will become larger at low temperature, we can not neglect it in this case. In addition to this TSOT, there exist other torques, i.e., the STT, we show it in Fig.4. The other physical observables such as charge current and thermal current are also demonstrated in the fig.1 and Fig2, respectively.

It should be pointed out that we only chose a simple uniform magnetization in our calculation, in fact the magnetization usually varies with position in some systems, especially in the domain wall. However, we can not solve the SBE (3) and (4) by the method of Fourier transformation when the magnetization is position dependent, it will be troubled by the convolution problem, we should deal with it by other complicated methods, that is left for future exploration.

## Acknowledgments

This study are supported by the National Key R&D Program of China (Grant No. 2018FYA0305804), and the Key Research Program of the Chinese Academy of Sciences (Grant No. XDPB08-3).